# Attosecond precision multi-km laser-microwave network

**Ming Xin,[1†*] Kemal Şafak,[1,2†] Michael Y. Peng,[3†] Aram Kalaydzhyan,[1] Wenting Wang,[1] Oliver D. Mücke,[1] and Franz X. Kärtner[1,2,3*]**

[1]Center for Free-Electron Laser Science, Deutsches Elektronen-Synchrotron, Notkestrasse 85, Hamburg 22607, Germany

[2]Department of Physics, University of Hamburg and the Hamburg Center for Ultrafast Imaging, Luruper Chaussee 149, 22761 Hamburg, Germany

[3]Research Laboratory of Electronics, Massachusetts Institute of Technology, Cambridge, Massachusetts 02139, USA

[†]These authors contributed equally to this work.

[*]email: ming.xin@cfel.de, franz.kaertner@cfel.de

**Abstract:**

**Synchronous laser-microwave networks delivering attosecond timing precision are highly desirable in many advanced applications, such as geodesy, very-long-baseline interferometry, high-precision navigation and multi-telescope arrays. In particular, rapidly expanding photon science facilities like X-ray free-electron lasers and intense laser beamlines require system-wide attosecond-level synchronization of dozens of optical and microwave signals up to kilometer distances. Once equipped with such precision, these facilities will initiate radically new science by shedding light on molecular and atomic processes happening on the attosecond timescale, such as intramolecular charge transfer, Auger processes and their impact on X-ray imaging. Here, we present for the first time a complete synchronous laser-microwave network with attosecond precision, which is achieved through new metrological devices and careful balancing of fiber nonlinearities and fundamental noise contributions. We demonstrate timing stabilization of a 4.7-km fiber network and remote optical-optical synchronization across a 3.5-km fiber link with an overall timing jitter of 580 and 680 attoseconds RMS, respectively, for over 40 hours. Ultimately we realize a complete laser-microwave network with 950-attosecond timing jitter for 18 hours. This work can enable next-generation attosecond photon-science facilities to revolutionize many research fields from structural biology to material science and chemistry to fundamental physics.**

## Introduction

The quest for isolated attosecond hard-X-ray pulses has dramatically intensified over the last few years[1] with the first observation of intramolecular charge transfer[2] and the discovery of ultrafast Auger processes altering the chemistry of matter on an attosecond-timescale[3,4]. Next-generation photon science facilities such as X-ray free-electron lasers (XFELs) (e.g., the European XFEL[5], FERMI[6], SwissFEL, LCLS[7] and LCLS II[8]) and intense laser beamline facilities (e.g., the Extreme Light Infrastructure ELI[9]) are emerging world-wide with the goal of generating sub-fs X-ray pulses with unprecedented brightness to capture ultrafast chemical and physical phenomena with sub-atomic spatio-temporal resolution[10,11]. These facilities, however, cannot fulfill this long-standing scientific dream without a high-precision timing distribution system. As illustrated in Fig. 1, the critical task of timing distribution is to synchronize various microwave and optical sub-sources across multi-km distances required for seeded FELs and attosecond pump-probe experiments. So far, there have been no reports demonstrating a synchronous laser-microwave network that permits attosecond precision across such distances. Hence, attosecond precision synchronization is the major obstacle that prevents attosecond-resolution photon science at hard-X-ray wavelengths.

Two basic timing distribution schemes have been reported to date. The first scheme uses traditional microwave signal distribution via amplitude modulation on a continuous-wave (cw) optical carrier[12]. This scheme solely depends on electronic phase-locking techniques and so far has not delivered better than ~100-fs root mean square (RMS) jitter facility-wide[13] due to low timing discrimination with microwaves and high noise floor at photo detection. The second scheme[14,15], which is pursued in this paper, uses ultralow-noise pulses generated by a mode-locked laser[16,17] as the timing signal to synchronize optical and microwave sources using balanced optical cross-correlators (BOC)[18] and balanced optical-microwave phase detectors (BOMPD)[19], respectively. In contrast to techniques used in frequency metrology[20-24], this approach eliminates the need for additional laser frequency combs at each end station since it utilizes the ultrashort optical pulses directly as time markers for precision timing measurements and features orders-of-magnitude higher timing stability. While this pulsed scheme has breached the 10-fs precision level[25-27], realizing and maintaining attosecond precision requires new metrological devices and better physical understanding of optical pulse shaping in fiber transmission and its impact on optical/microwave measurements at the fundamental level. This advanced level of physical/technical comprehension is a prerequisite to unfold the full potential of next-generation attosecond photon science facilities.

To this matter, we have thoroughly analyzed pulse propagation effects in the fiber link and systematically eliminated noise limitations in the whole system to develop a new pulsed timing distribution system. Here, we present the first demonstration of a laser-microwave network with attosecond timing precision, which corresponds to a >10× improvement in timing stability compared to the previously published results[27,28] and satisfies the imperative and challenging synchronization requirements for next-generation photon science facilities.

## Materials and method

### Simulation model

we developed a numerical model to simulate pulse timing jitter during nonlinear pulse propagation in the fiber link. In this model, the master equation of a fast saturable absorber mode-locked laser is solved using the fourth-order Runge–Kutta in the interaction picture (RK4IP) method[29]. Laser timing jitter is generated by adding amplified spontaneous emission noise during each iteration of RK4IP, whose amount corresponds to the measured jitter of the master laser[26]. The pulse train is centered at 1550 nm with 170-fs pulse width and 216-MHz repetition rate. Self-phase modulation, self-steepening and Raman effect are considered in the link. Both the nonlinear Schrödinger equation for the link transmission and the pulse coupled field equation for second-harmonic generation in the BOC are solved using the split-step Fourier method with adaptive step length. The BOC characteristic (i.e., the BOC output voltages with respect to the initial delay of the two input pulses) is calculated for each round trip link pulse against a new laser reference pulse. The timing offset of the zero-crossing position in the BOC characteristic is identified as timing error.

To calculate link-enhanced excess jitter (in Fig. 2a and 2b), the simulation is repeated for a train of laser pulses in the presence of pulse timing jitter. The RMS of the timing errors from all the BOC characteristics is calculated to obtain the overall link-enhanced excess jitter. To calculate the power fluctuation-induced drift (Fig. 2c and 2d), only one pulse is simulated in the absence of pulse timing jitter, since this source of timing error is deterministic.

**BOC characteristics**

Two methods are used to experimentally characterize the timing sensitivity of a BOC. The first method is for the case where the two input pulse trains in the BOC have the exact same repetition rate. The relative delay between the pulse trains is swept with a motorized delay stage, while the response voltage of the BOC is recorded with a data acquisition card. The slope of the measured BOC characteristic at its zero-crossing is the timing sensitivity. In the second case, two laser input pulse trains with different repetition rates are combined in a BOC, a train of BOC characteristics is generated at the BOC output. One can simultaneously record a BOC characteristic on an oscilloscope and measure the instantaneous repetition-frequency difference (RFD) between the lasers with photodetectors and electrical mixers. The real time scale of the BOC characteristic can be calibrated by multiplying the oscilloscope time scale with the ratio of the RFD to the laser repetition rate. Coarse frequency tuning is performed in advance to ensure a small RFD so that the BOC characteristic is not limited by the balanced photo detector (BPD) bandwidth.

**BOMPD characteristics**

The phase sensitivity of the BOMPD can similarly be measured using a free-running laser and a microwave oscillator. Due to aliasing during electro-optic sampling, the effective frequency difference between the oscillators is $f_{\mathrm{beat}} = f_{\mathrm{RF}} \bmod f_{\mathrm{rep}}$, where $f_{\mathrm{RF}}$ is the frequency of the radio frequency (RF) oscillator and $f_{\mathrm{rep}}$ is the fundamental repetition rate of the laser. The BOMPD output voltage signal will be a train of BOMPD response characteristics with a repetition frequency equal to this frequency difference $f_{\mathrm{beat}}$. One can record this frequency difference and a single BOMPD characteristic simultaneously with an oscilloscope. The oscilloscope time scale multiplied by the angular frequency difference ($2\pi f_{\mathrm{beat}}$) represents the phase error between the optical and RF signal relative to the RF signal frequency. The BOMPD phase sensitivity is defined as the slope of the BOMPD characteristic at its zero-crossing in units of V/rad. The timing sensitivity in units of V/s is obtained by further multiplying the phase sensitivity by the RF oscillator angular frequency ($2\pi f_{RF}$).

### Other measurement methods

The noise floors of all BOCs in the experiments are limited by the detector electronic noise, since the signal power from second-harmonic generation is relatively low. The feedback timing precision of the BOC is calculated as the integrated RMS noise voltage of the BPD within the locking bandwidth, calibrated by the BOC timing sensitivity.

The long-term drift data in Fig. 4e are measured by filtering the out-of-loop signal with a 1-Hz lowpass anti-alias filter and recording with a data acquisition card at a 2-Hz sampling rate. The jitter spectral density data in Fig. 4g are direct baseband power spectrum measurement of the BOC/BOMPD output on a signal source analyzer. In Fig. 4h, the data above 1 Hz is the integration of the timing jitter spectrum in Fig. 4g; the data below 1 Hz is integrated using the Fourier transform of the drift data in Fig. 4e.

## Results and discussion

### Link-induced timing jitter and drift

In the pulsed timing distribution approach (Fig. 1), an optical pulse train with ultralow jitter (timing signal) is generated from a mode-locked laser (master laser), and distributed through polarization-maintaining dispersion-slope-compensated fiber (PM-DCF) links in a star network topology. At the end of each link, an output coupler partially reflects the timing signal back towards the link input. The timing offset between the returning link pulse and a new pulse from the master laser is measured with a BOC. The error voltage signal from the BOC is fed back to a variable delay line in the link path to compensate for any detected timing errors. Using this feedback scheme, various environmental fluctuations, including mechanical stress, acoustics and temperature imposed on the link can be significantly corrected for. The fundamental limits to this noise suppression scheme are set by the inherent laser noise, BOC detection noise floor, reference path noise, and link-induced noise. Of these limitations, the link-induced noise will dominate and prevent optimum link performance if not properly accounted for.

Based on our numerical model, residual link dispersion and nonlinearities add considerable excess jitter in the high-frequency range above 1 kHz even in the absence of environmental noise. First, pulse center-frequency fluctuations are coupled to timing jitter via residual second-order dispersion (SOD) and third-order dispersion (TOD) (Fig. 2a). This jitter contribution, often called Gordon-Haus jitter[30], can amount to 0.1 and 0.3 fs for uncompensated SOD equivalent to 2 and 3 m of standard PM fiber, respectively. Second, spontaneous emission noise is coupled to timing jitter and its impact is further enhanced by link nonlinearities (Fig. 2b). This jitter is bounded at 0.13 fs for average power levels below +12 dBm (corresponding pulse peak power $P_{peak}$=430 W) but escalates to 1.4 fs at +14 dBm ($P_{peak}$=682 W). Since these excess jitter contributions can transfer to the link output through the feedback loop (see Supplementary Fig. S2d), fiber-link dispersion and nonlinearities must be minimized to achieve attosecond link stability.

Moreover, link power fluctuations on slower timescales can similarly introduce timing errors that degrade link stability. Long-range compensation for link stabilization is performed by a free-space motorized delay line (MDL) with long delay arms; e.g., a 10-cm range is required to correct for ±1.5-K temperature change in a 3.5-km link. Movement of the delay stage introduces inevitable beam misalignments that cause link power fluctuations. These fluctuations induce temporal shifts in the pulse center-of-gravity through a composite effect of residual SOD, TOD and nonlinearity (see Supplementary Eq. S28). Although a center-of-gravity shift appears as a deterministic shift in the zero-crossing position of the in-loop BOC characteristic, the link stabilization feedback will unknowingly track this shift and introduce it into the link path, causing a timing error at the link output. Simulations are performed using typical values observed in the experiment. Fig. 2c shows that residual TOD can induce timing errors up to 5 fs for +8-dBm link power with ±5% fluctuations. Fig. 2d indicates that +10-dBm link power is the threshold before significant amplitude-to-timing conversion occurs due to severe nonlinear pulse distortions and may result in 4 fs of timing error from ±5% power fluctuations. Link power variations and residual TOD must be minimized to achieve long-term attosecond precision.

**Laser-microwave network**

Taking the outcomes of this jitter analysis into account, an attosecond-precision laser-microwave network is demonstrated using the setup in Fig. 3a. The timing signal from the master laser is distributed through a network that contains two independent fiber links of 1.2-km and 3.5-km length operated in parallel. The link outputs are used to synchronize a remote laser (e.g., serving as a pump-probe laser at the FEL end station in Fig. 1) and a voltage-controlled oscillator (VCO) (e.g., serving as a microwave reference of the FEL linear accelerator in Fig. 1) simultaneously.

A beta-barium borate (BBO) crystal with large birefringence is used in each locking BOC to realize a polarization-noise-suppressed BOC (PNS-BOC) for improved noise performance, as shown in Fig. 3b. At the PNS-BOC output, since there are no time-dependent error voltages introduced by undesired pulse components ($E_{Ls}$ and $E_{Rp}$), each PNS-BOC can be locked exactly at the zero-crossing position of its BOC characteristic. This is crucial to make the BOC itself perfectly independent with laser amplitude noise so as to achieve attosecond precision. A feedback precision of ~2 as for the laser locking PNS-BOC is achieved with a low-noise BPD.

Residual SOD and TOD of links are compensated with additional dispersion-compensating fiber to suppress the link-induced Gordon-Haus jitter and to minimize the output pulse durations for high signal-to-noise ratio (SNR) in the BOCs. The link power is adjusted to minimize the nonlinearity-induced jitter as well as to maximize the SNR for BOC locking. To eliminate power fluctuations caused by beam misalignment in the MDL, a feedback signal is sent to the EDFA to control its pump current (Fig. 3c).

In Fig. 3d, free-space-coupled BOMPD (FSC-BOMPD) is developed and employed for optical-to-microwave locking. The free-space components at the optical input can efficiently reduce long-term drifts caused by the environment, and the delay stages can enable precise phase tuning without backlash, microwave reflection and loss. Compared with other optical-microwave phase detectors[31,32], this new device is unaffected by optical input power fluctuations and can provide high SNR and a >10× improvement in terms of long-term timing stability simultaneously (see Supplementary Eqs. (S30-37)), which are essential to achieve attosecond precision in the laser-microwave network.

Three characterization setups are adopted (Fig. 3e): two timing link monitoring signals (TLM1 and 2) are sent to an out-of-loop BOC to evaluate the link network performance;

the master laser monitoring signal (MLM) and the remote laser output signal (RLO) are sent to another BOC to characterize the remote laser synchronization; finally, the remote microwave output (RMO) and RLO are compared with an out-of-loop FSC-BOMPD. The third setup is of great significance since it directly measures the true relative timing jitter between a remotely-synchronized mode-locked laser and a microwave source, which has never been shown before.

The timing sensitivity of the link locking PNS-BOC 1 and 2, laser locking PNS-BOC and VCO locking FSC-BOMPD are 1 mV/fs, 2 mV/fs, 7.9 mV/fs and 0.25 mV/fs respectively (Fig. 4a-d), which are large enough to support tight locking for the laser-microwave network. Stabilization of the 4.7-km link network is operated continuously for 52 hours. The residual timing drift between TLM 1 and 2 below 1 Hz is only 200 as RMS (Fig. 4e, red curve); the relative timing drift instability is $6\times10^{-17}$ with 1 s averaging time $\tau$ and reduced to $7.3\times10^{-21}$ at $\tau = 10^4$ s (Fig. 4f, red circle). The equivalent phase noise at 10.833 GHz is lower than -110 dBc/Hz at 1 Hz and goes below -145 dBc/Hz after 20 kHz (Fig. 4g, red curve); whereas the total integrated timing jitter from 6 μHz to 1 MHz is only 580 as (Fig. 4h, red curve). Remote laser synchronization is achieved successfully for over 44 hours without interruption. Residual timing drift is less than 100 as RMS (Fig. 4e, blue curve), which is an order-of-magnitude improvement over previous results[27], and corresponds to a relative timing instability of $2.5\times10^{-22}$ in 50000 s (Fig. 4f, blue triangle). The integrated jitter is only 200 as in the range of 7 μHz – 1 kHz and 680 as for 7 μHz – 1 MHz (Fig. 4h, blue curve). Finally, the whole laser-microwave network shows an unprecedented long-term precision of 670 as RMS out-of-loop drift over 18 hours (Fig. 4e, black curve). Compared with previous frequency-comb-based microwave transfer results[28], this setup includes an additional fiber link and a remote laser synchronization system, yet it still achieves more than an order-of-magnitude improvement. The relative timing stability between the two remote-synchronized devices within the full frequency range from 15 μHz to 1 MHz is only 950 as RMS (Fig. 4h, black curve). To the best of our knowledge, this is the first attosecond-precision demonstration of remote optical-to-microwave synchronization as well as the first demonstration of a synchronous laser-microwave network.

Based on the feedback model in the Supplementary Information, the out-of-loop jitter is contributed by the environmental noise imposed on the link, the electronic noise of the system, the master laser's inherent jitter and the link-induced jitter (see Supplementary Eq. (S2)). In our experiment, most of the environmental noise is below 1 kHz and can be well suppressed by the feedback loop. The link-induced jitter is also minimized by choosing the minimum link operating power required for tight link/laser/microwave locking. Therefore, the bumps from 1 kHz to 20 kHz of all three curves in Fig. 4g are mainly attributed to the master laser's inherent jitter and the system electronic noise, which may even be amplified at those resonant frequencies of the feedback loop if not paid attention (see Supplementary Fig. S2b and c). For the laser-microwave network results in Fig. 4h (black curve), the power line noise at 50 Hz and its harmonics, contribute about 250 as jitter, which can be removed by using cleaner power supplies. The residual drift below 100 mHz, is limited by the length fluctuations of the conventional coaxial cables in all RF paths of the FSC-BOMPDs, which can be improved by reducing all electronics into an integrated board or using special phase-stable cables with a much lower thermal-expansion ratio.

## Conclusion

In summary, by adopting new metrological timing detectors PNS-BOCs and FSC-BOMPDs, and reducing link-induced timing jitter and drift from nonlinear pulse propagation effects, long-term-stable attosecond timing precision has been achieved across a 4.7-km fiber link network between remote optical and microwave devices. The attosecond-precision laser-microwave network will enable next-generation FELs and other science facilities to operate with the foreseen timing precision to unfold their full potential. This will drive new scientific efforts towards the making of atomic and molecular movies at the attosecond timescale, thereby opening up many new research areas in biology, drug development, chemistry, fundamental physics and material science. Besides, this technique will also accelerate developments in many other fields requiring high spatio-temporal resolution such as ultrastable clocks[33,34], gravitational wave detection[35] and coherent optical antenna arrays[36].

## Acknowledgements

The authors thank Qing Zhang for supporting laboratory construction, Guoqing Chang for valuable discussions on simulations, Stefano Valente for initial link simulations, Patrick T. Callahan for discussions on link stabilization and Wei Liu for the help on laser-microwave network characterization experiments. The authors acknowledge financial support by the European Research Council under the European Union's Seventh Framework Program (FP/2007-2013) / ERC Grant Agreement n. 609920 and the Cluster of Excellence: The Hamburg Centre for Ultrafast Imaging-Structure, Dynamics and Control of Matter at the Atomic Scale of the Deutsche Forschungsgemeinschaft.

## Author contributions

F.X.K. and O.D.M. initiated the project. M.X. did the jitter limitation analysis and simulation. M.X., K.S. and M.Y.P. contributed with the fiber network stabilization and optical-optical synchronization system. M.Y.P., A.K. and M.X. designed the FSC-BOMPD. M.X., K.S., A.K. and W.W realized the laser-microwave network. All authors prepared the manuscript.

## Competing financial interests

The authors declare no competing financial interests.

## Figure Legends

**Figure 1 Layout of a timing distribution system for next generation free-electron lasers.** A single timing stabilized link length is up to 3.5 km, which corresponds to the length of the European XFEL[5], the largest photon science facility in the world close to completion.

**Figure 2 Analysis of link-induced timing jitter and drift.** A 216-MHz, 170-fs pulse train from a mode-locked laser, with an integrated timing jitter of 0.3 fs above 1 kHz[26], is sent

into a 3.5-km PM fiber link with an erbium-doped optical fiber amplifier (EDFA) and a fiber mirror near the link output. Relative timing jitter between the round-trip pulse train and the original pulse train is calculated in **(a)** and **(b)** for frequencies above 1 kHz. Timing drift errors introduced by the link stabilization feedback are shown in **(c)** and **(d)**, which are related to environmental fluctuations occurring on slow timescales below 10 Hz. **(a)** Link-induced Gordon-Haus jitter due to residual link dispersion. B2: the link's residual SOD normalized by the SOD of 1-m standard polarization maintaining fiber PM 1550; B3: the link's residual TOD normalized by the TOD of 1-m PM 1550 fiber. The ripples along the B3 axis are mainly due to pulse shaping induced by phase fluctuations through TOD. **(b)** Link-enhanced timing jitter due to fiber nonlinearities and the corresponding BOC characteristics at each input power level. The timing sensitivity increases with input power. This jitter caused by fiber nonlinearity needs to be carefully minimized in practice because it easily reaches fs-level before a visible distortion of the BOC characteristic can be observed. **(c)** Timing drift induced by link input power fluctuations for different B3 values; for each curve, the input power level is +8 dBm and B2=-0.13. **(d)** Timing drift induced by link input power fluctuations for different input power levels; for each curve, B2=-0.13 and B3=18.7.

**Figure 3 Experimental setup. (a)** Schematic of the laser-microwave network. **(b)** All BOCs consist of a single 4-mm-long periodically-poled $KTiOPO_4$ (PPKTP) crystal operating in a double-pass configuration with appropriate dichroic beam splitter and mirror (DBS, DM). Ideally, the input pulses ($E_{Lp}$ and $E_{Rs}$) are aligned along the two principal axes of the PPKTP crystal for maximum second-harmonic generation. Due to finite polarization extinction ratios in the optical elements upstream from BOCs, there will be undesired pulse

components ($E_{Ls}$ and $E_{Rp}$) projected along the undesired polarization axes. In PNS-BOC, a linear material with large birefringence is put before the BOC. This material adds a significant delay to the erroneous pulses such that they do not overlap and interfere with the second-harmonic generation process in PPKTP. In our setup, a BBO crystal is used to provide the required birefringence, whose cut angle is carefully selected to make sure that it cannot generate any nonlinear process. **(c)** In the timing link stabilization, the output of the link locking PNS-BOC is sent to an MDL, a fiber stretcher (FS) and a PM-EDFA to compensate the long-term temperature drift, fast jitter imposed on the link and link power fluctuations, respectively. **(d)** Inside the FSC-BOMPD, a self-referenced signal ($0.5Mf_R$, $f_R$ is the repetition rate of the optical input signal) is derived from the input optical pulse train to bias the Sagnac interferometer (SGI) at quadrature. The pulse train in the main SGI path performs electro-optical sampling of the RF input signal ($Nf_R$) to convert phase error into amplitude modulation. The modulated pulse train is detected and down-converted in-phase using another self-referenced signal ($0.5f_R$) to baseband. The voltage phase error signal can be filtered by a proportional-integral (PI) controller and fed back to the microwave source for optical-to-microwave synchronization (Alternatively, a feedback signal can also be applied to the laser of the optical input for microwave-to-optical synchronization). The zero-crossings of the microwave signal are then phase-locked to the pulse positions of the pulse train (i.e., $\Delta\theta$=0). (**e**) Out-of-loop characterization setups.

**Figure 4 Measurement results of three characterization setups.** The characteristics of link locking PNS-BOC 1 and 2, laser locking PNS-BOC and VCO locking FSC-BOMPD are shown in **(a)**, **(b)**, **(c)** and **(d)**, respectively. **(e)** Long-term timing drift (sampling rate=2 Hz). **(f)** Timing drift instability (Allan deviation) versus averaging time $\tau$. A fit of $\tau^{-1}$ slope

is shown. **(g)** Timing jitter spectral density and the corresponding phase noise referenced to 10.833 GHz at [1 Hz, 1 MHz]. **(h)** Integrated timing jitter.

# Figure

## Figure 1

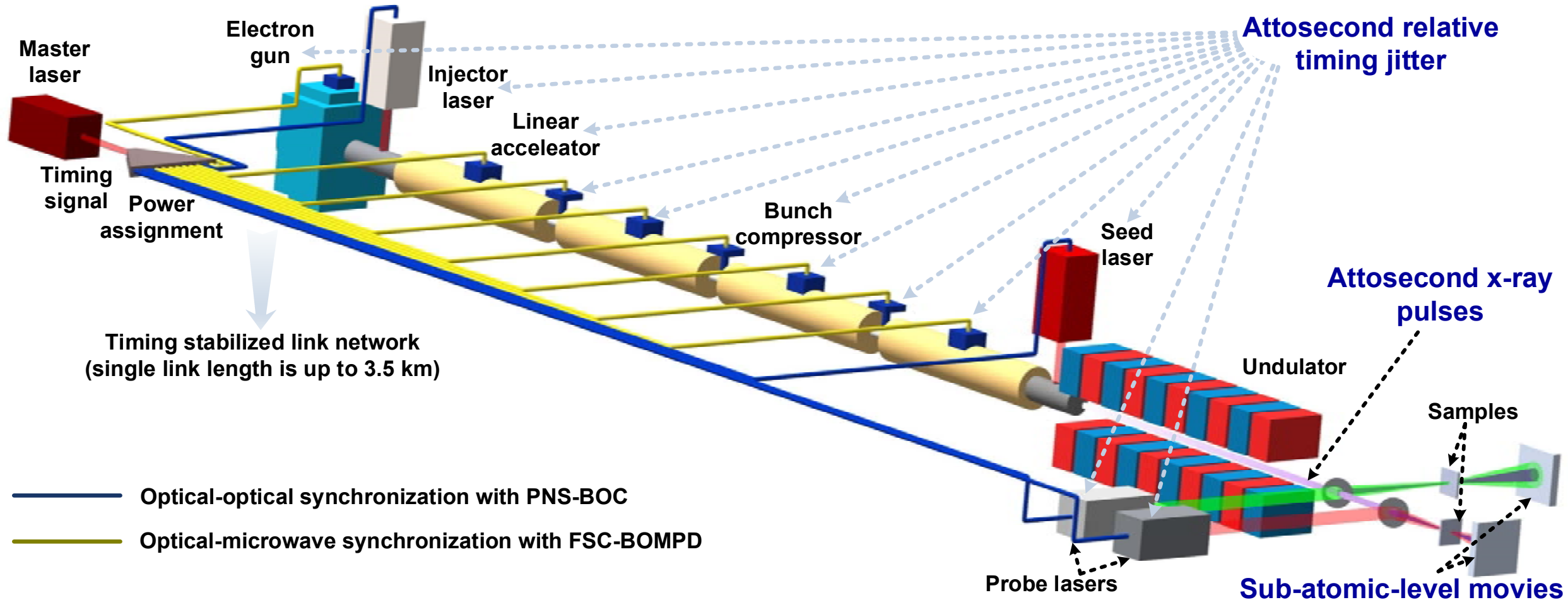

**Figure 2**

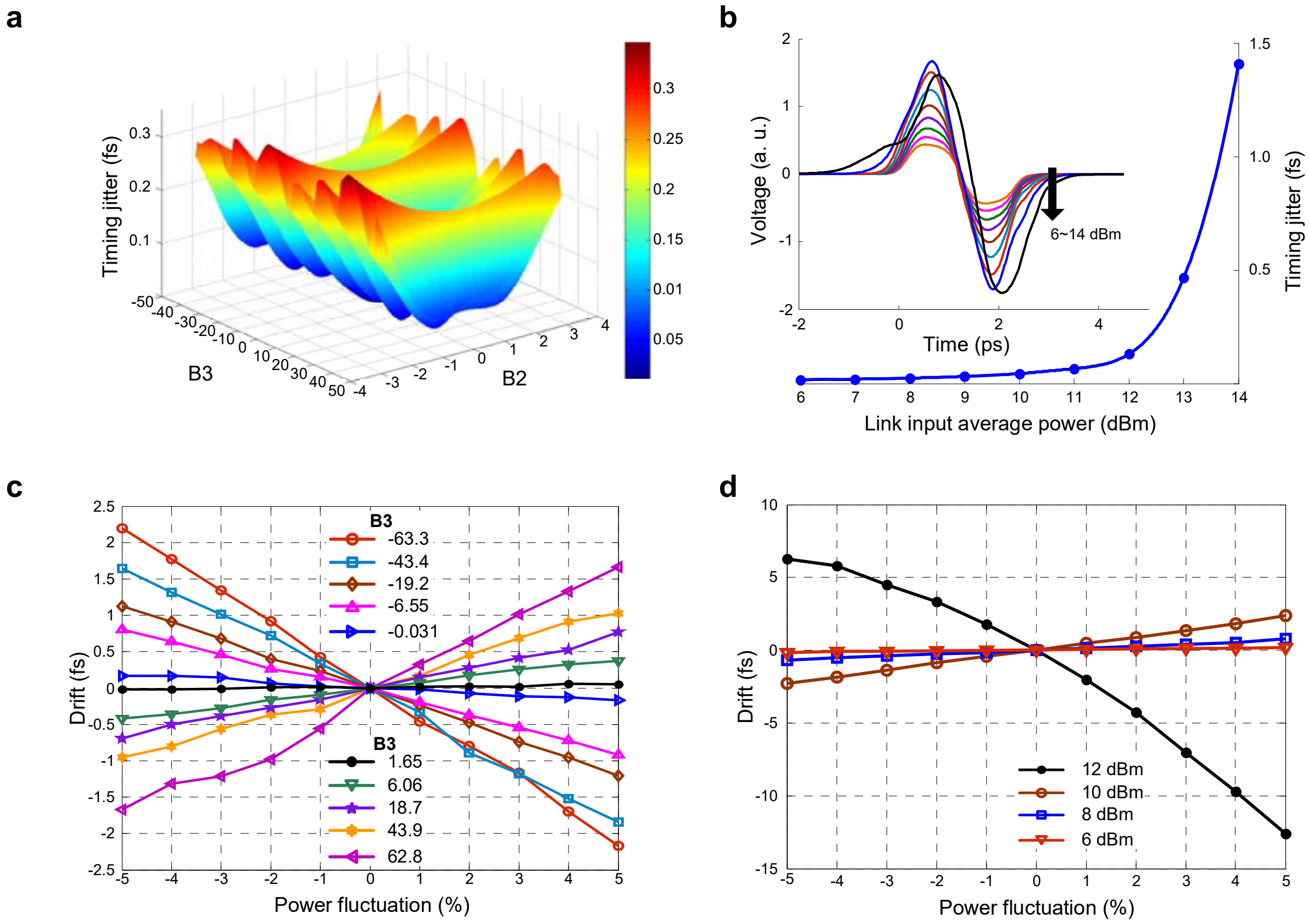

# Figure 3

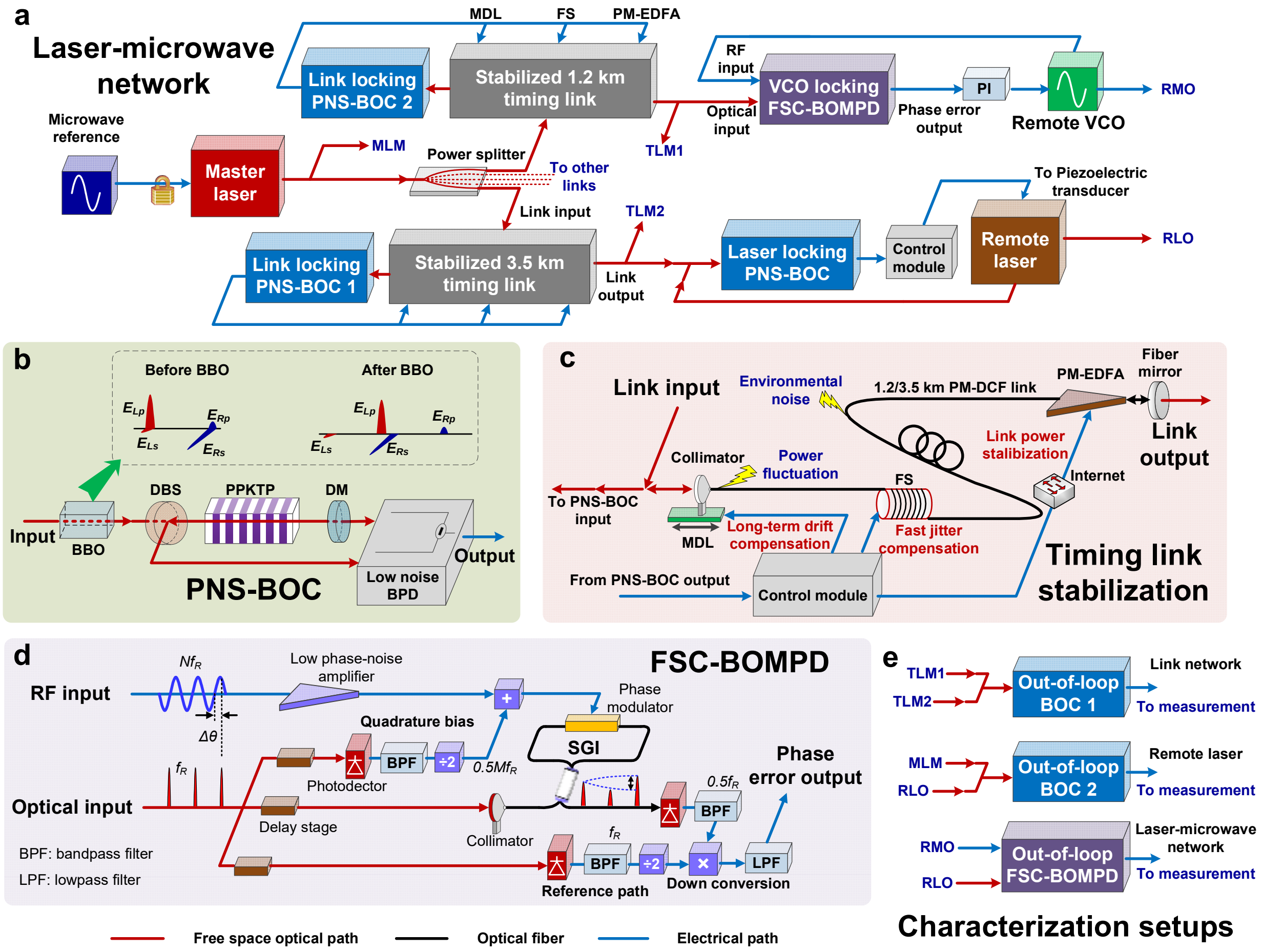

# Figure 4

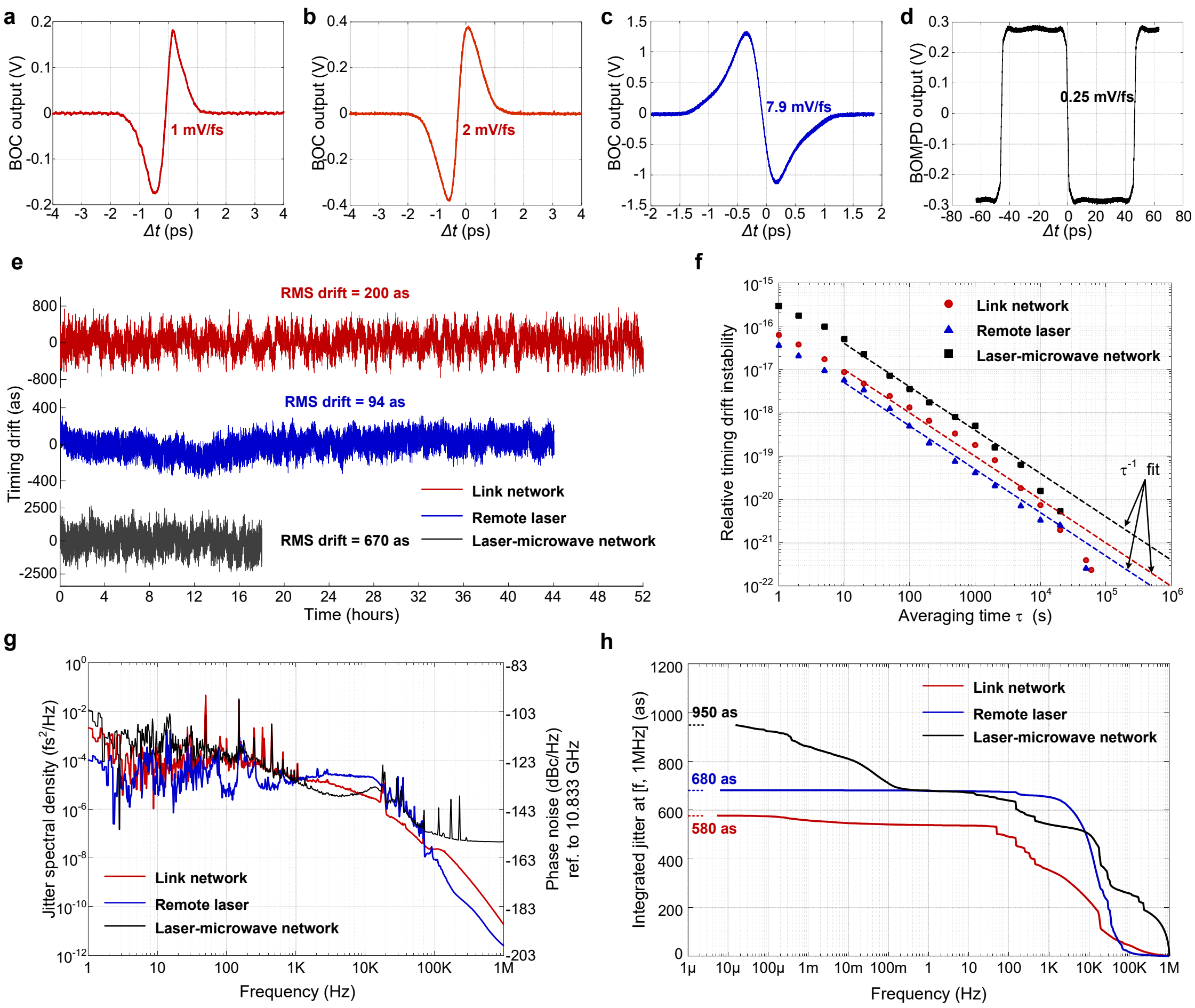

a
b
c
d
BOC output (V)
BOMPD output (V)
Δt (ps)
1 mV/fs
2 mV/fs
7.9 mV/fs
0.25 mV/fs
e
RMS drift = 200 as
RMS drift = 94 as
RMS drift = 670 as
Timing drift (as)
Time (hours)
Link network
Remote laser
Laser-microwave network
f
Relative timing drift instability
Averaging time τ (s)
τ^-1 fit
g
Jitter spectral density (fs²/Hz)
Phase noise (dBc/Hz) ref. to 10.833 GHz
Frequency (Hz)
h
Integrated jitter at [f, 1MHz] (as)
950 as
680 as
580 as

**Supplementary Information**

# Attosecond precision multi-km laser-microwave network

**Ming Xin,[1*] Kemal Şafak,[1,2] Michael Y. Peng,[3] Aram Kalaydzhyan,[1] Wenting Wang,[1] Oliver D. Mücke,[1] and Franz X. Kärtner[1,2,3,*]**

[1]*Center for Free-Electron Laser Science, Deutsches Elektronen-Synchrotron, Notkestrasse 85, Hamburg 22607, Germany*
[2]*Department of Physics, University of Hamburg and the Hamburg Center for Ultrafast Imaging, Luruper Chaussee 149, 22761 Hamburg, Germany*
[3]*Research Laboratory of Electronics, Massachusetts Institute of Technology, Cambridge, Massachusetts 02139, USA*
[*]Email: ming.xin@cfel.de, franz.kaertner@cfel.de

Here, we would like to provide the reader with more detailed information on the research presented in the main paper. First, the simulations investigating noise sources induced by the fiber link will be discussed. Then, the details of the fiber link network, remote laser synchronization and laser-microwave network experiments will be explained together with that of the FSC-BOMPD.

## I. Simulation of link-induced noise

To investigate the link-induced noise, we have developed a feedback and numerical model of the timing link to simulate pulse timing jitter during nonlinear pulse propagation in the fiber link. The results show that even in the absence of environmental noise, residual link dispersion and nonlinearities add considerable excess jitter.

### Feedback model of the timing link

Fig. S1 shows a flow diagram for the timing link feedback model. In the in-loop section, the detected timing jitter $Y_{IL}$ by the in-loop BOC is first converted to voltage by the transfer function $H_B$, amplified by $H_G$ with electronic noise $E_B$, and fed to the PI controller $H_{PI}$ in a negative feedback configuration. The PI output, along with additive electronic noise $E_{PI}$, are amplified and converted back to timing delays $J_C$ by $H_{PZT}$. Due to the round-trip propagation, $J_C$ is added to the link delay twice. Furthermore, the round-trip link transmission time delay $2T_L$ needs to be considered for both $J_C$ and master laser inherent jitter $J_I$. During link propagation, additional jitter from link-induced Gordon-Haus jitter[S1] and link-enhanced timing jitter due to fiber nonlinearities arises through $H_{LR}$. In practice, the environmental noise is usually significant below 1 kHz; therefore, the forward and backward link transmissions for a 3.5 km link impose the same environmental jitter value $J_E$ on the link pulses. The in-loop jitter $Y_{IL}$ is the relative jitter between the round-trip link pulses and new pulses from the master laser. The out-of-loop jitter $Y_{OL}$ is the relative jitter between the link output pulses and new pulses from the master laser. For the out-of-loop measurement, $J_C$ and $J_I$ experience a single-pass link propagation delay $T_L$. Furthermore, $J_I$ experiences additional link-induced jitter via $H_{LF}$, and environmental jitter is contributed by $J_E$.

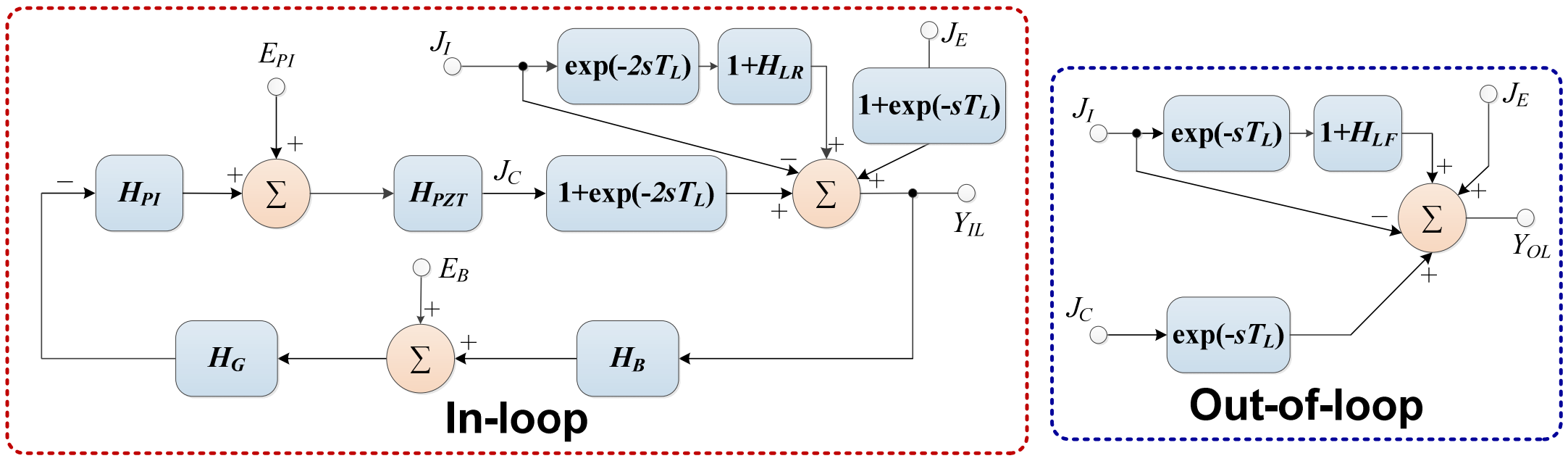

**Figure S1.** Feedback model of the timing link ($J_I$, inherent jitter of the master laser; $J_E$, integrated environmental jitter imposed on the link for single-trip link transmission; $Y_{IL}$, detected timing jitter by the in-loop BOC; $H_B$, transfer function of the in-loop BOC; $E_B$, electronic noise of the BPD in the in-loop BOC; $H_G$, transfer function of the amplifier in the BPD; $H_{PI}$, transfer function of the PI controller; $E_{PI}$, electronic noise of the PI controller; $H_{PZT}$, transfer function of the fiber stretcher and its driver; $J_C$, equivalent timing delay generated by the control loop for compensation; $T_L$, single-trip link transmission time; $H_{LR}$, equivalent transfer function of the link-induced time jitter for round-trip transmission; $H_{LF}$, equivalent transfer function of the link-induced timing jitter for forward link transmission; $Y_{OL}$, relative timing jitter between the link output pulses and the original pulses from the master laser).

Based on this model, we have

$$\begin{aligned} Y_{IL} &= J_I\left[\exp(-2sT_L)\left(1+H_{LR}\right)-1\right]+J_C\left[1+\exp(-2sT_L)\right]+J_E\left[1+\exp(-sT_L)\right] \\ J_C &= \left[-(Y_{IL}H_B+E_B)G_BH_{PI}+E_{PI}\right]H_{PZT} \\ Y_{OL} &= J_I\left[\exp(-sT_L)\left(1+H_{LF}\right)-1\right]+J_C\exp(-sT_L)+J_E \end{aligned} \tag{S1}$$

Then we obtain

$$Y_{OL} = C_EJ_E + C_NJ_N + C_IJ_I + C_LH_{LR}J_I \tag{S2}$$

where

$$
\begin{aligned}
&J_N = -\frac{E_B}{H_B} + \frac{E_{PI}}{H_B G_B H_{PI}} \\
&C_E = \frac{1+\mathbf{H}\left[1-\exp(-sT_L)\right]}{1+\mathbf{H}\left[1+\exp(-2sT_L)\right]} \\
&C_N = \frac{\mathbf{H}\exp(-sT_L)}{1+\mathbf{H}\left[1+\exp(-2sT_L)\right]} \\
&C_I = \frac{1+\mathbf{H}\left[1-\exp(-sT_L)\right]}{1+\mathbf{H}\left[1+\exp(-2sT_L)\right]}\left[\exp(-sT_L)-1\right] \\
&C_L = \frac{k+\mathbf{H}\left[k-\exp(-2sT_L)(1-k)\right]}{1+\mathbf{H}\left[1+\exp(-2sT_L)\right]}\exp(-sT_L) \\
&\mathbf{H} = H_B G_B H_{PI} H_{PZT} \\
&k = \frac{H_{LF} J_I}{H_{LR} J_I}
\end{aligned}
\tag{S3}
$$

As Eq. (S2) indicates, the out-of-loop jitter $Y_{OL}$ has four main contributions: the environmental noise imposed on the link, the electronic noise of the system, the master laser's inherent jitter and the link-induced jitter, with coefficients $C_E$, $C_N$, $C_I$ and $C_L$, respectively, where $k$ is a variable in the range [0, 1] that represents the degree of symmetry of link-induced jitter between the forward and backward link propagations. It should be noted that in our model, the environmental noise is treated as a discrete effect with a delay $T_L$ between forward and backward trip. A more accurate approach is to divide the link into n-segments, and inject the environmental noise along different parts of the link, which will change $C_E$ slightly.

The coefficients $C_i$, $i$=$E$, $N$, $I$, $L$ for a 3.5-km link can be calculated using the transfer functions of the experimental equipment. Fig. S2a-c show the calculated coefficients $|C_E|$, $|C_N|$, $|C_I|$, for five different PI controller gain settings. High gain is necessary to efficiently suppress the environmental noise below 1 kHz (Fig. S2a). However, the electronic noise from the BPD in the BOC and the PI controller rises with increasing gain (Fig. S2b). Furthermore, large gain peaks appear in Fig. S2a-c at frequencies $n/4T_L$ (n=1, 3, 5…) as well as around the resonant frequency of the fiber stretcher (about 18 kHz), if the feedback gain is too high. Therefore, to optimize the system performance, a medium gain needs to be adopted – see black curves in Fig. S2a-c. With this gain setting, $|C_I|$ in Fig. S2c exponentially increases from 0.02 at 1 kHz to 4.6 at about 16 kHz, which means that the master laser's inherent jitter can appear in the out-of-loop measurement through the feedback loop.

Also using such moderate gain value, $|C_L|$ with different $k$ values is calculated in Fig. S2d. If the link-induced jitter from the forward and backward link transmission is almost identical, like in the case of the Gordon-Haus jitter in Fig. 2a, then $k$ is about 0.5, and $|C_L|$ is increased from 0.12 at 1 kHz to 2 at 14.6 kHz (Fig. S2d). In the case of the nonlinearity-induced jitter shown in Fig. 2b, if the backward link transmission power is much higher than for the forward path, the majority of the jitter is coming from the backward trip, $k$ is almost 0. In Fig. S2d, $|C_L|$ is about 0.5 at 1 kHz and approach 1.5 at 16 kHz. On the other hand, if the forward power is greater than for the backward path, $k$ is close to 1, and $|C_L|$ increases from 0.5 to 2.4 at [1 kHz, 14 kHz]. Above all, the link-induced jitter in Fig. 2a and b can transfer to the link-output through the feedback loop, either partially, completely or even with amplification.

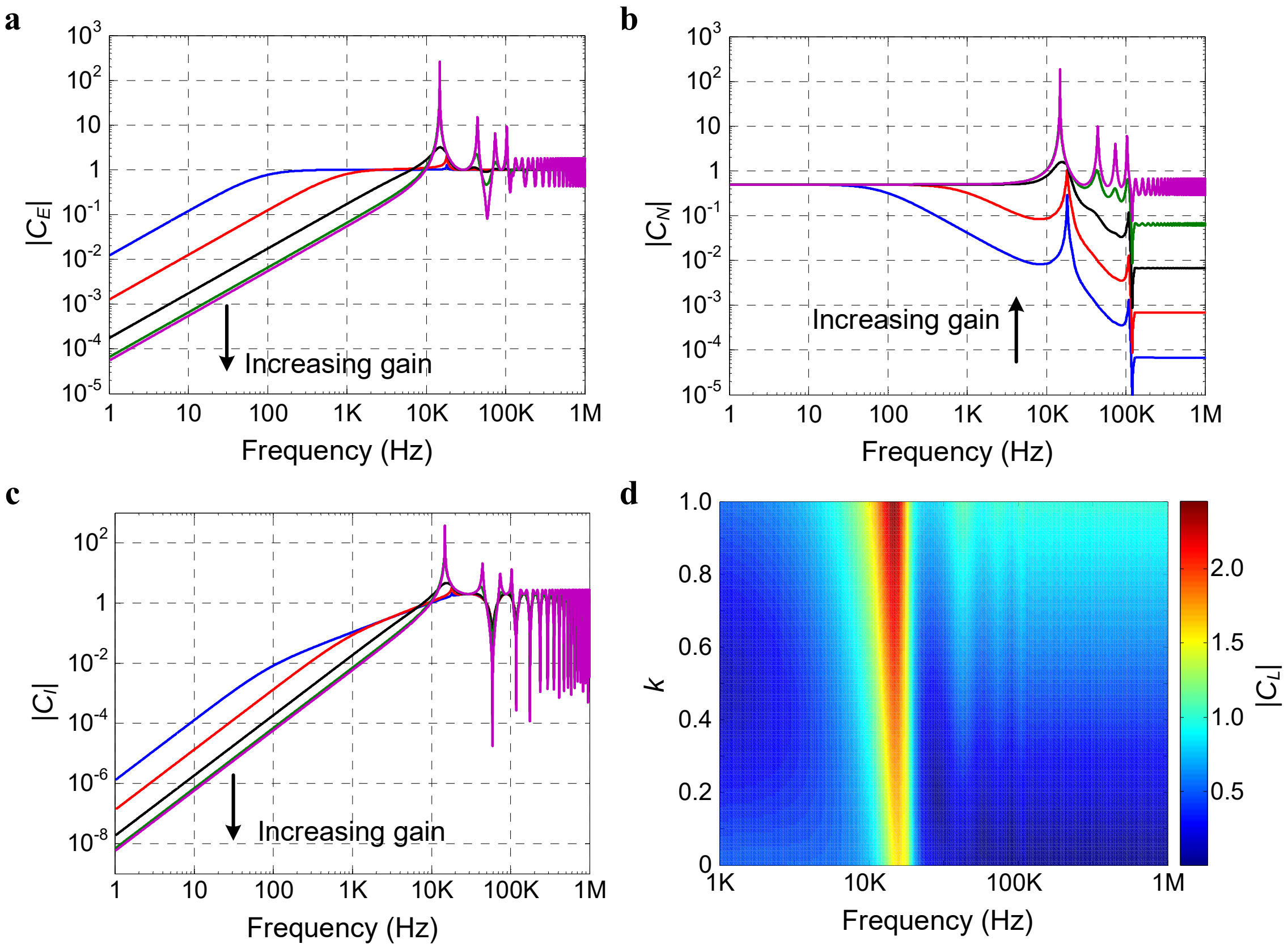

**Figure S2.** Simulation results for a 3.5-km timing link feedback model. **(a)** the coefficient for the environmental noise imposed on the link; **(b)** the coefficient for the electronic noise; **(c)** the coefficient for the master laser's inherent jitter; **(d)** the coefficient for the link-induced jitter with different $k$ values.

## Timing drift induced by link power fluctuations

An analytical model is developed to support the simulation results in Fig. 2c and 2d. The sum-frequency generation process in a BOC in undepleted-fundamental-frequency approximation is governed by the coupled wave equations[S2]:

$$\left[\frac{\partial}{\partial z}+\left(\frac{1}{v_1}-\frac{1}{v_3}\right)\frac{\partial}{\partial t}\right]E_1=0$$
$$\left[\frac{\partial}{\partial z}+\left(\frac{1}{v_2}-\frac{1}{v_3}\right)\frac{\partial}{\partial t}\right]E_2=0 \tag{S4}$$
$$\frac{\partial}{\partial z}E_3=i\frac{d_{eff}\omega_3}{n_3c}E_1E_2$$

where $E_1$, $E_2$ and $E_3$ are the electric fields of the link-reflected pulse, reference pulse and sum-frequency generation pulse, respectively, in the retarded time frame of $E_3$ with $v_i$ denoting the group velocities of $E_i$ ($i$=1, 2, 3). $d_{eff}$ is the nonlinear optical coefficient, $\omega_3$ is the carrier angular frequency of $E_3$, $n_3$ is the refractive index of $E_3$ in the crystal, and $c$ is the vacuum speed of light. Pulse shaping due to dispersion in the crystal is not included. Based on Eq. (S4), we have

$$E_3(t)=K\int_0^{L_C}E_1(t-k_1z)E_2(t-k_2z)dz \tag{S5}$$

where $L_C$ is the crystal length and

$$K=i\frac{d_{eff}\omega_3}{n_3c}$$
$$k_1=\frac{1}{v_1}-\frac{1}{v_3} \tag{S6}$$
$$k_2=\frac{1}{v_2}-\frac{1}{v_3}$$

If there is an initial delay $T_D$ between $E_1$ and $E_2$, and $L_C$ is sufficiently long, then Eq. (S5) can be approximated by

$$E_3(t,T_D)=K\int_0^{\infty}E_1(t-k_1z)E_2(t-k_2z-T_D)dz \tag{S7}$$

Suppose the reference pulse is so short that $E_2$ can be approximated by a Dirac delta function, then Eq. (S7) can be simplified as

$$E_3(t,T_D)=\begin{cases}KE_1(t-k_1\frac{t-T_D}{k_2}), & t>T_D\\ 0, & \text{else}\end{cases} \tag{S8}$$

The forward direction sum-frequency generation power of the BOC is

$$P_F(T_D) = \int_{-\infty}^{+\infty} \left|E_3(t, T_D)\right|^2 dt \tag{S9}$$

Based on Eqs. (S8) and (S9), we have

$$P_F(T_D) = \left|K\right|^2 \frac{k_2 - k_1}{k_2} \int_{T_D}^{\infty} \left|E_1(t)\right|^2 dt \tag{S10}$$

In first-order linear approximation, Eq. (S10) can be simplified to

$$P_F(T_D) = \begin{cases} \left|K\right|^2 \dfrac{k_2 - k_1}{k_2} E, & T_D < T_{D0} - \dfrac{1}{2} T_{BOC} \\ \left|K\right|^2 \dfrac{k_1 - k_2}{k_2} \dfrac{E}{T_{BOC}} \left( T_D - T_{D0} - \dfrac{1}{2} T_{BOC} \right), & T_{D0} - \dfrac{1}{2} T_{BOC} \le T_D \le T_{D0} + \dfrac{1}{2} T_{BOC} \\ 0, & T_D > T_{D0} + \dfrac{1}{2} T_{BOC} \end{cases} \tag{S11}$$

where $T_{D0}$ is the zero-crossing time of the BOC, $T_{BOC}$ is the linearly varying range of $P_F$, and

$$E = \int_{-\infty}^{+\infty} \left|E_1(t)\right|^2 dt \tag{S12}$$

The temporal center-of-gravity of $E_1$ is defined as the first order moment of $t$,

$$T_{COG} = \frac{1}{E} \int_{-\infty}^{+\infty} t \left|E_1(t)\right|^2 dt = \frac{1}{E} \lim_{T \to \infty} \int_{-T}^{T} t \left|E_1(t)\right|^2 dt \tag{S13}$$

Since

$$\begin{aligned} \int_{-T}^{T} t \left|E_1(t)\right|^2 dt &= t \int_{-\infty}^{t} \left|E_1(\tau)\right|^2 d\tau \Big|_{-T}^{T} - \int_{-\infty}^{+\infty} \int_{-\infty}^{t} \left|E_1(\tau)\right|^2 d\tau dt \\ &= -TE + \frac{k_2}{\left|K\right|^2 (k_2 - k_1)} \int_{-T}^{T} P_F(t) dt \end{aligned} \tag{S14}$$

Based on (S11), (S13) and (S14), we have

$$T_{COG} = T_{D0} \tag{S15}$$

Usually the pulse shape of $E_1$ is distorted during the link transmission. Now let us transfer to the retarded time frame of pulse propagation in the fiber link. Define $A(z, T)$ as the slowly varying amplitude of the link pulse envelope. $A(z, T)$ is governed by the nonlinear Schrödinger equation[S3]:

$$\frac{\partial A}{\partial z}=\left(-\frac{\alpha}{2}-\frac{i\beta_2}{2}\frac{\partial^2}{\partial T^2}+\frac{\beta_3}{6}\frac{\partial^3}{\partial T^3}\right)A(z,T)+i\gamma\left(|A|^2 A+\frac{i}{\omega_0}\frac{\partial}{\partial T}(|A|^2 A)-T_R A\frac{\partial |A|^2}{\partial T}\right) \quad \text{(S16)}$$

where $\alpha$ is the fiber loss, $\beta_2$ is the second-order dispersion (SOD) coefficient, $\beta_3$ is the third-order dispersion (TOD) coefficient, $\gamma$ is the nonlinear parameter for self-phase modulation (SPM), $\omega_0$ is the carrier angular frequency, and $T_R$ is the Raman parameter. The timing link consists of two fiber sections: standard polarization maintaining (PM) fiber, and PM dispersion compensating fiber (DCF). Here it is assumed that $\alpha$ and $T_R$ are constant and $\beta_2$, $\beta_3$ and $\gamma$ are $z$-dependent.

The slowly varying amplitude of the pulse envelope can be written in terms of a normalized amplitude function $U$ on a time scale $t$ normalized to the input pulse width $T_0$ as

$$A(z,T)=\sqrt{P_0}\exp(-\alpha z/2)U(z,t) \quad \text{(S17)}$$

$$t=\frac{T}{T_0} \quad \text{(S18)}$$

where $P_0$ is the peak power of the input pulse. Combining Eqs. (S16-18), $U(z,t)$ satisfies

$$U_z=-\frac{i\beta_2}{2T_0^2}U_{tt}+\frac{\beta_3}{6T_0^3}U_{ttt}+i\gamma P_0\exp(-\alpha z)\left(|U|^2 U+\frac{i}{\omega_0 T_0}(|U|^2 U)_t-\frac{T_R}{T_0}U|U|_t^2\right) \quad \text{(S19)}$$

where the subscripts $t$ and $z$ indicate partial derivatives, e.g., $U_z \equiv \partial U/\partial z$, $U_{tt} \equiv \partial^2 U/\partial t^2$.

The center-of-gravity of $A(z, T)$ can be written as

$$t_{COG}=\frac{\int_{-\infty}^{+\infty} T|A(T)|^2\,dT}{\int_{-\infty}^{+\infty}|A(T)|^2\,dT}=T_0\int_{-\infty}^{+\infty} t|U(t)|^2\,dt \quad \text{(S20)}$$

Based on Eq. (S15), in first-order linear approximation, the timing drift detected by the BOC can be characterized by changes in $t_{COG}$.

Using Eq. (S20) and (S19), we obtain

$$\frac{dt_{COG}}{dz} = T_0 \int_{-\infty}^{+\infty} t \left[ \frac{i\beta_2}{2T_0^2}\left(U_{tt}^* U - U_{tt} U^*\right) + \frac{\beta_3}{6T_0^3}\left(U_{ttt}^* U + U_{ttt} U^*\right) - \frac{3\gamma P_0 \exp(-\alpha z)}{\omega_0 T_0} \left|U\right|_t^2 \left|U\right|^2 \right] dt \quad \text{(S21)}$$

At time positions far from a single pulse (t→±∞), it is reasonable to assume that the pulse temporal profile decays to zero and is well-behaved:

$$\begin{aligned} &\lim_{t\to\pm\infty} U(z,t) = 0 \\ &\lim_{t\to\pm\infty} \frac{\partial^n U}{\partial t^n} = 0, \quad n = 1,2,... \end{aligned} \quad \text{(S22)}$$

Utilizing Eq. (S22), Eq. (S21) can be rewritten as

$$\frac{dt_{COG}}{dz} = \frac{i\beta_2}{2T_0}\Omega + \frac{\beta_3}{2T_0^2}\Gamma + \frac{3\gamma P_0 \exp(-\alpha z)}{2\omega_0} S \quad \text{(S23)}$$

where

$$\begin{aligned} &\Omega = \int_{-\infty}^{+\infty} (U_t U^* - U U_t^*) dt \\ &\Gamma = \int_{-\infty}^{+\infty} \left|U_t\right|^2 dt \\ &\mathrm{S} = \int_{-\infty}^{+\infty} \left|U\right|^4 dt \end{aligned} \quad \text{(S24)}$$

The derivative of Ω and Γ with respect to $z$ can be further calculated as follows

$$\begin{aligned} &\frac{d\Omega}{dz} = i2\gamma P_0 \exp(-\alpha z) \frac{T_R}{T_0} F_1(z) \\ &\frac{d\Gamma}{dz} = i\gamma P_0 \exp(-\alpha z) \left( F_2(z) - \frac{T_R}{T_0} F_3(z) \right) \end{aligned} \quad \text{(S25)}$$

where

$$\begin{aligned} &F_1(z) = \int_{-\infty}^{+\infty} \left( \left|U\right|_t^2 \right)^2 dt \\ &F_2(z) = \int_{-\infty}^{+\infty} \left|U\right|_t^2 (U U_t^* - U^* U_t) dt \\ &F_3(z) = \int_{-\infty}^{+\infty} \left|U\right|_{tt}^2 (U U_t^* - U^* U_t) dt \end{aligned} \quad \text{(S26)}$$

With numerical calculation, $F_2(z)$ and $F_3(z)$ can be easily verified to be nonzero for Gaussian and hyperbolic secant pulses with nonzero residual dispersion.

Substituting the integrated form of (S25) into (S23) and integrating (S23) over the link length $L$ yields

$$t_{COG} = \frac{1}{2T_0^2}\int_{-\infty}^{+\infty}|U_t(t,0)|^2\,dt\int_0^L \beta_3 dz \\ -P_0\left[\begin{array}{l} 2\frac{T_R}{T_0^2}\int_0^L \beta_2\int_0^z \gamma\exp(-\alpha z_1)F_1(z_1)dz_1dz + \frac{1}{T_0^2}\int_0^L \beta_3\int_0^z \gamma\exp(-\alpha z_1)F_2(z_1)dz_1dz \\ -\frac{T_R}{T_0^3}\int_0^L \beta_3\int_0^z \gamma\exp(-\alpha z_1)F_2(z_1)dz_1dz + \frac{3}{2\omega_0}\int_0^L \gamma\exp(-\alpha z)S(z)dz \end{array}\right] \quad \text{(S27)}$$

Suppose that the input power with fluctuation $P_0(1+\delta)$ corresponds to a center-of-gravity with fluctuation $t_{COG}+\Delta t_{COG}$ for a given power fluctuation ratio δ. According to Eq. (S27), this yields

$$\frac{\Delta t_{COG}}{\delta} = -P_0\left[\begin{array}{l} 2\frac{T_R}{T_0^2}\int_0^L \beta_2\int_0^z \gamma\exp(-\alpha z_1)F_1(z_1)dz_1dz + \frac{1}{T_0^2}\int_0^L \beta_3\int_0^z \gamma\exp(-\alpha z_1)F_2(z_1)dz_1dz \\ -\frac{T_R}{T_0^3}\int_0^L \beta_3\int_0^z \gamma\exp(-\alpha z_1)F_3(z_1)dz_1dz + \frac{3}{2\omega_0}\int_0^L \gamma\exp(-\alpha z)S(z)dz \end{array}\right] \quad \text{(S28)}$$

The first term in the brackets on the right side of Eq. (S28) is dependent on the Raman response and SOD, whereas the second term is related to SPM and TOD. Furthermore, the third term corresponds to the Raman response and TOD, and the last term indicates the self-steepening effect. Hence, it can be concluded that the drift induced by power fluctuations is proportional to the input power level and is a combined effect of residual SOD, TOD, SPM, self-steepening and Raman response.

## II. Fiber link network stabilization

The master laser used for the laser-microwave network operates with 1554.7-nm central wavelength, 170-fs pulse duration and 216.6675-MHz repetition rate, which is locked to a microwave reference with a 10-Hz feedback bandwidth to suppress long-term drifts. In practice, the laser power can be divided into several simultaneously operating timing links for the remote synchronization at different locations. Here, the laser power is split into two timing links. The input signal to each link stabilization setup is further divided into reference and link path pulses. The reference path lengths are set as short as possible (4 cm) to minimize timing errors introduced by environmental noise.

To assure that the forward and backward link transmission accumulates the same amount of jitter, the link pulse must travel along the same polarization axis during round-trip propagation. Therefore, a 45° Faraday rotator before the fiber link is necessary to introduce a 90° round-trip polarization rotation to direct the reflected link pulse towards the BOC in the link stabilization block.

The BPDs of the link locking PNS-BOCs have two 1-MHz monitor outputs and a balanced output port with a 3-dB bandwidth of 100 MHz, while the BPD output bandwidth of the out-of-loop BOC 1 is 1 MHz.

Both links are constructed with a section of standard PM 1550 fiber, followed by a section of PM dispersion compensating fiber. The ratios of dispersion coefficients between the two fiber types are

$$\begin{aligned} \frac{\beta_{2d}}{\beta_{2p}} &= -5.663 \\ \frac{\beta_{3d}}{\beta_{3p}} &= -5.613 \end{aligned} \tag{S29}$$

where $\beta_{2p}$ and $\beta_{3p}$ are the SOD and TOD coefficient of PM 1550, and $\beta_{2d}$ and $\beta_{3d}$ are the SOD and TOD coefficient of PM dispersion compensating fiber. Due to this ratio difference, it is not possible to compensate SOD and TOD simultaneously. For example, if the SOD of the 3.5-km link is completely eliminated, 26-m worth of TOD from PM 1550 would be uncompensated. An optical auto-correlator is used to confirm the temporal duration of the link pulses. Due to the residual TOD, pulse durations of ~400 fs are obtained for both links. Based on Fig. 2c, this residual TOD can introduce a drift of at least 2 fs for ±5% link power fluctuations. Link power stabilization in Fig. 3c is therefore indispensable to achieve attosecond drift.

A detailed schematic of the link network characterization setup is given in Fig. S3. Two polarizers are placed before and after the link to improve the polarization extinction ratio. The output voltage of the PI controller is divided into two paths. The first path compensates for fast noise in the link and consists of a high-voltage amplifier that drives a PM fiber stretcher. The second path compensates for long-term environmental drift and consists of a data acquisition card (DAQ) card that samples the timing error and controls the MDL through a Labview program. Link power stabilization is achieved by measuring the BPD monitor ports in the PNS-BOCs and feeding back to the pump current of the EDFA through a TCP/IP network. While the timing drift due to power fluctuations in the forward link transmission can be neglected (because the link power is relatively low), the power stabilization scheme is necessary to remove the excess timing drift in the reverse link transmission after EDFA amplification for improved out-of-loop performance.

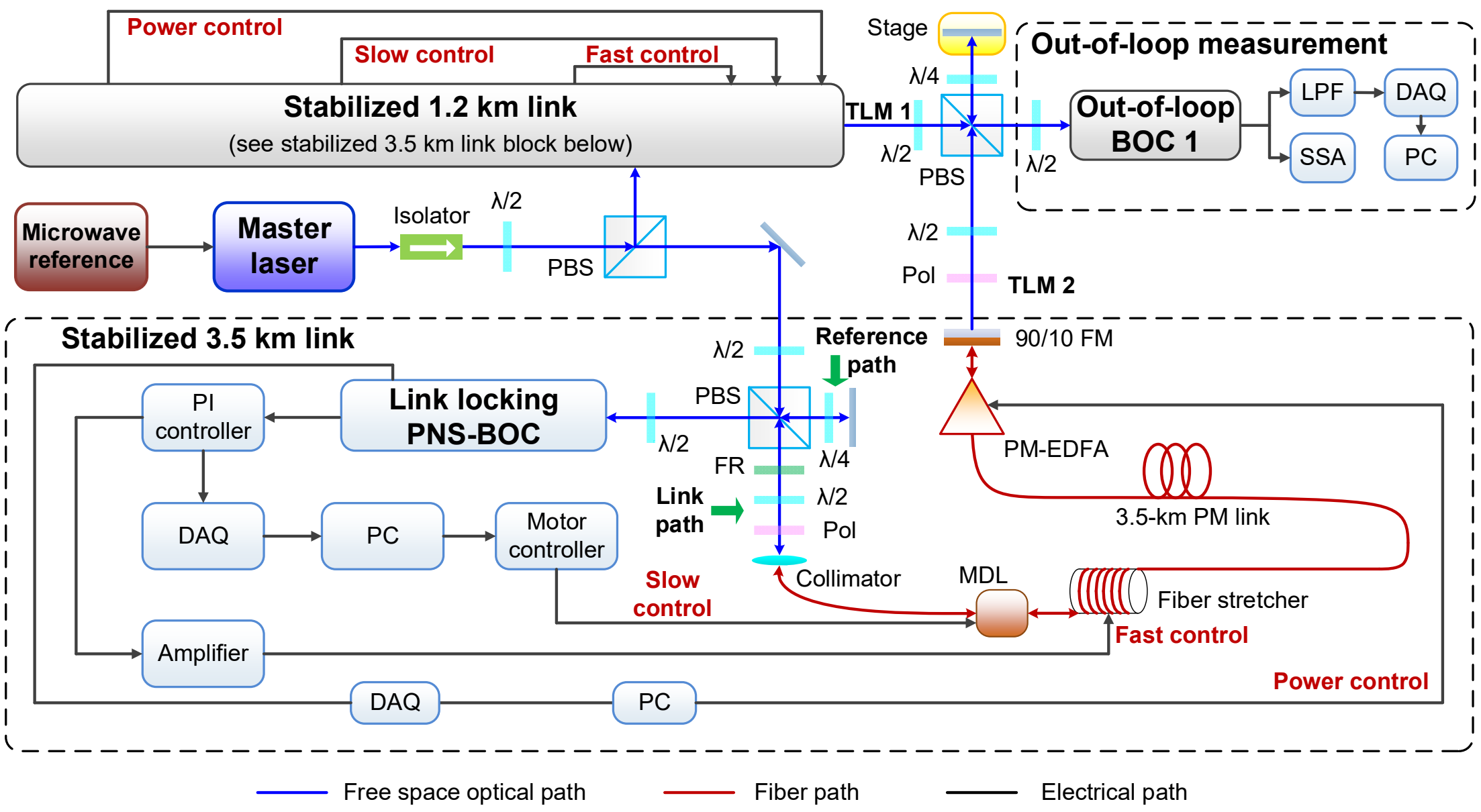

**Figure S3.** Detailed scheme of the link network stabilization (PBS, polarization beam splitter; λ/2, half-wave plate; λ/4, quarter-wave plate, FR, 45° Faraday rotator; Pol, polarizer; 90/10 FM, 90/10 transmission/reflection fiber mirror; LPF, 1-Hz low pass filter; SSA, signal source analyzer; PC, computer).

To minimize timing errors resulting from thermally-induced length fluctuations in the reference paths, all free-space optics are mounted on a temperature-stabilized breadboard with a Super-Invar surface sheet. With temperature fluctuations controlled below ±0.05 K, the effective timing-instability of free-space beam paths due to thermal expansion is ±1 as/cm. Lead foam is placed beneath the setup to damp table vibrations. A two-layer enclosure with acoustic heavy foil for the inner layer and high-density polyethylene (HDPE) for the outer layer is built to provide acoustic isolation for all optical components, except for the two fiber links which are placed outside of the enclosure and subjected to environmental changes.

Fig. S4 shows the monitored results during the link network stabilization demonstration. During 52 hours, about 8.6 ps and 35 ps link drift are compensated by the MDLs in the 1.2 km and 3.5 km long links, respectively. Temperature and relative humidity on the fiber links changed by 0.3 K and 4% respectively. With the help of power feedback control, power fluctuations in the 1.2-km link are suppressed to within ±0.2%. Due to the coarse resolution of the EDFA pump current, the power fluctuations in the 3.5-km link can only be stabilized to within ±1% and serves as the main drift contribution in Fig. 4h (red curve) from 300 μHz to 1 Hz.

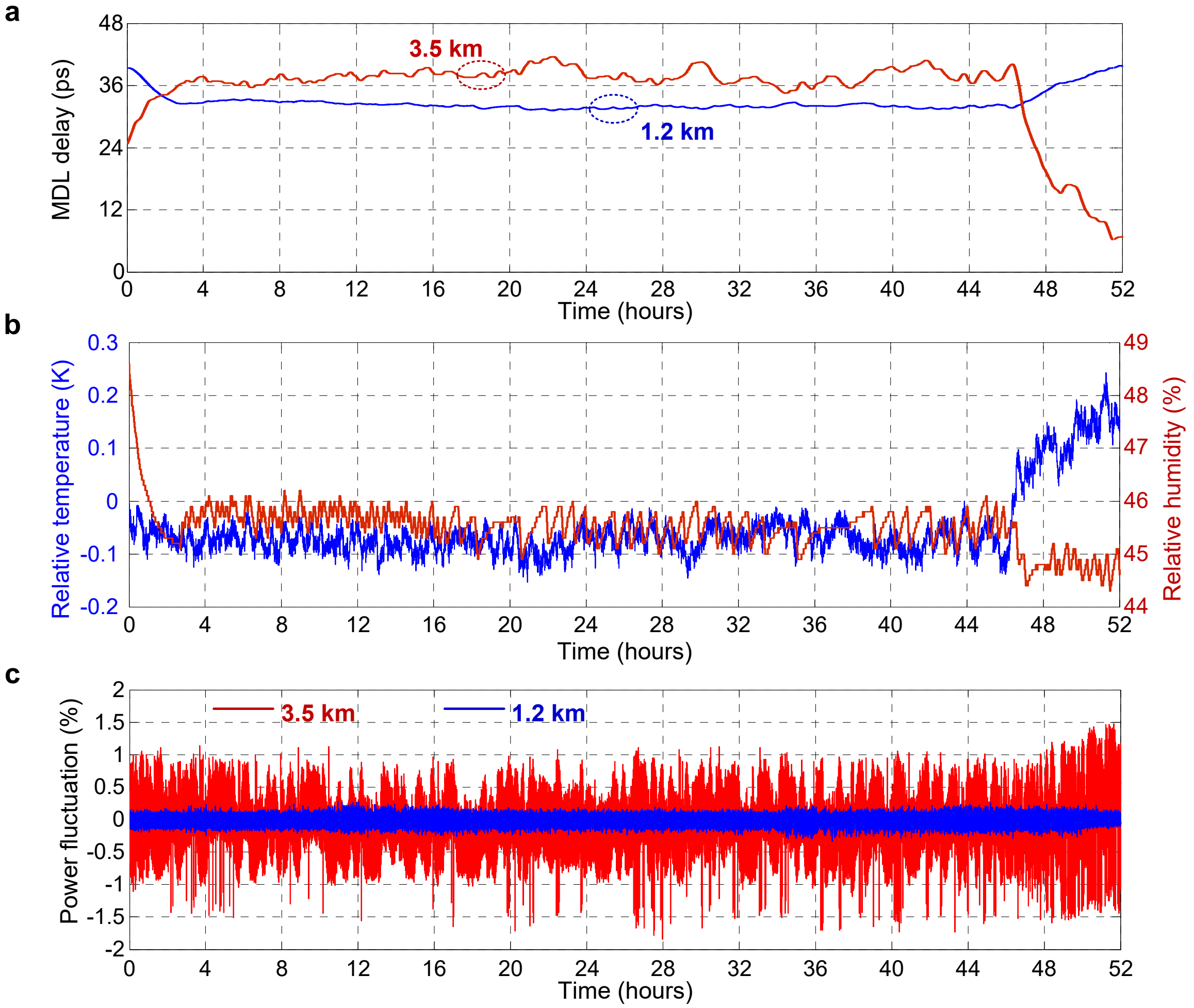

**Figure S4.** Monitored signals in a link network stabilization experiment. **(a)** Compensated link drift by the MDLs; **(b)** environmental temperature and humidity change; **(c)** power fluctuation after power control.

## III. Remote laser synchronization

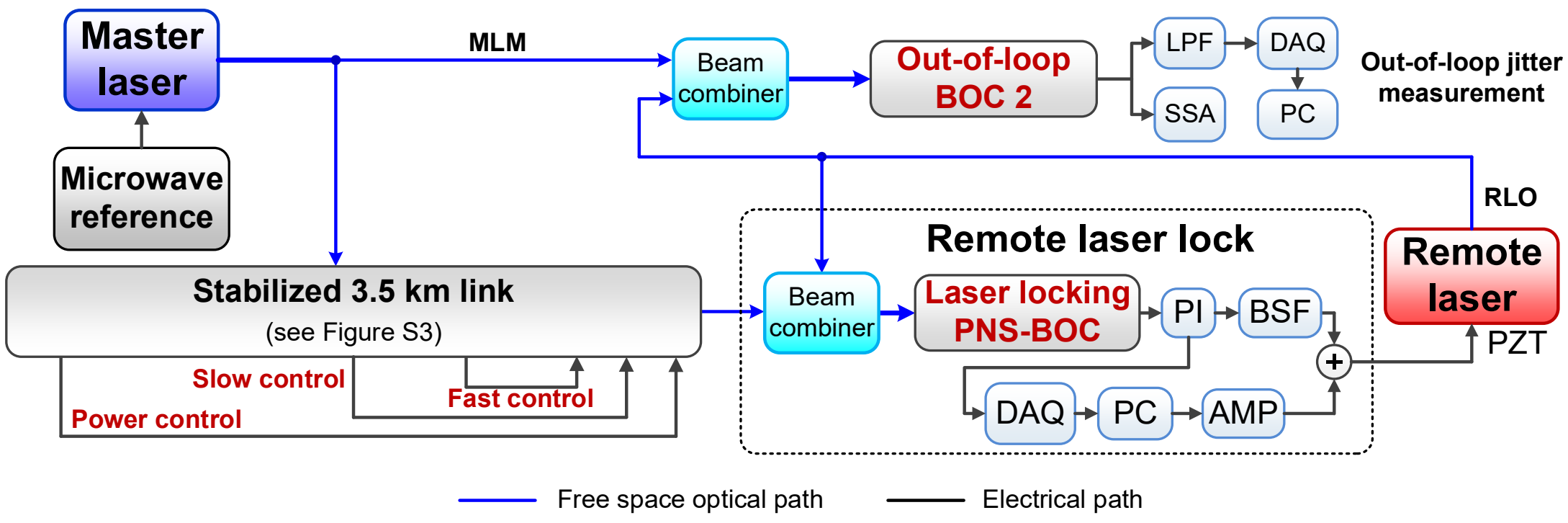

**Figure S5.** Detailed schematic of the remote laser synchronization (BSF, band stop filter; PZT, piezoelectric transducer).

In Fig. 3a, the remote laser operates at 1553.4-nm central wavelength with 170-fs pulse width and 216.6685-MHz repetition rate. Pulse trains from the 3.5 km link output and the remote laser are combined with orthogonal polarization and detected by the laser-locking PNS-BOC with a timing sensitivity of 7.9 mV/fs. As shown in Fig. S5, the laser-locking PNS-BOC output voltage is filtered by a PI controller. Due to the limited bandwidth of the amplifier, the PI output is separated into two paths to optimize feedback control for slow and fast noise independently. The first path is sampled by a data acquisition card and analyzed by a Labview program to generate a DC offset voltage to compensate slow timing drifts. A high-gain amplifier is used to extend the compensation range. The second path, which compensates fast noise, is not amplified to minimize phase excursions at high offset frequencies to achieve high locking bandwidth; a band stop filter is used to eliminate potential feedback loop resonances. The voltage summer output recombines the two paths to drive the piezoelectric transducer (PZT) of the remote laser, which has a sensitivity of 17.4 Hz/V.

During the 44-hour operation of remote laser synchronization (see Fig. S6), the link stabilization compensates about 61-ps link drift while the remote laser synchronization corrects for a 400-Hz repetition rate drift of the remote laser. The temperature variation is about 0.6 K and the relative humidity change is about 4% (measured on the fiber link), which are typical laboratory environmental conditions.

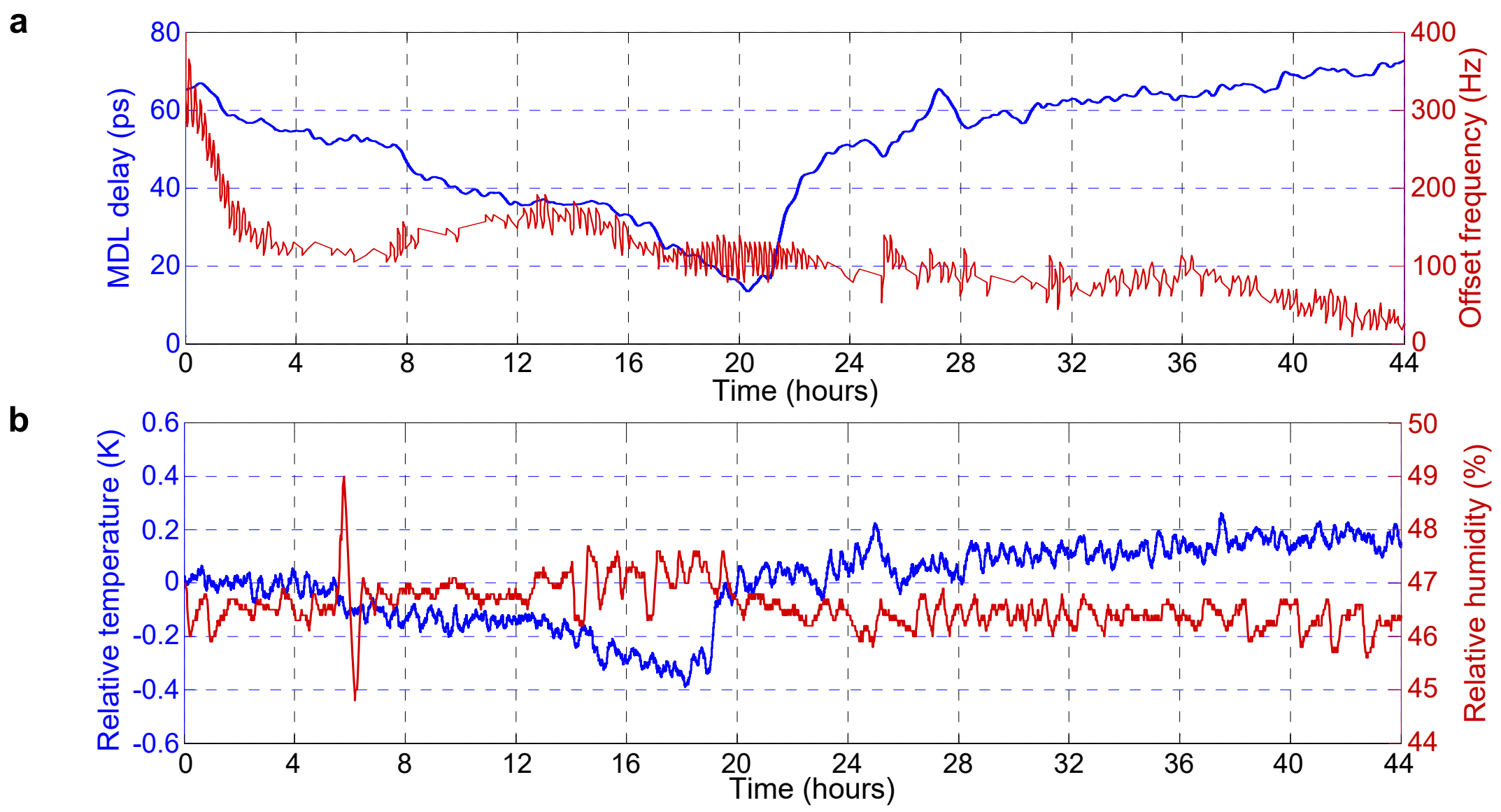

**Figure S6.** Monitoring results in a remote laser synchronization experiment. **(a)** Compensated drift by the MDL and the PZT offset frequency; **(b)** temperature and humidity change of the fiber link.

## IV. Free-space-coupled balanced optical-microwave phase detector

To improve the timing stability, we developed a new FSC-BOMPD, as shown in Fig. S7. Key characteristics in this BOMPD architecture are summarized as follows: first, free space components, such as PBS, half-wave plates and quarter-wave plates are used before the SGI, and the total length of the SGI is spliced to be as short as possible. This efficiently reduces the long-term drifts caused by the environment. Second, high-frequency (multi-GHz) modulation in the phase modulator is enforced to ensure unidirectional phase modulation. This eliminates the repetition-rate dependence of the SGI, thus improving its robustness and long-term stability. Third, error signal demodulation is performed at the lowest possible frequency to maximize SNR at photo detection and to minimize thermally induced phase changes in the RF signal paths for long-term stability. Fourth, free space delay stages are used to control the relative delay between different paths; this enables precise phase tuning without backlash, microwave reflection and loss when compared with RF phase shifters. Lastly, AM- and PM-sensitive signal paths are optimized to achieve an AM-PM suppression ratio of -50 dB, which is a high figure-of-merit for optical-to-microwave conversion.

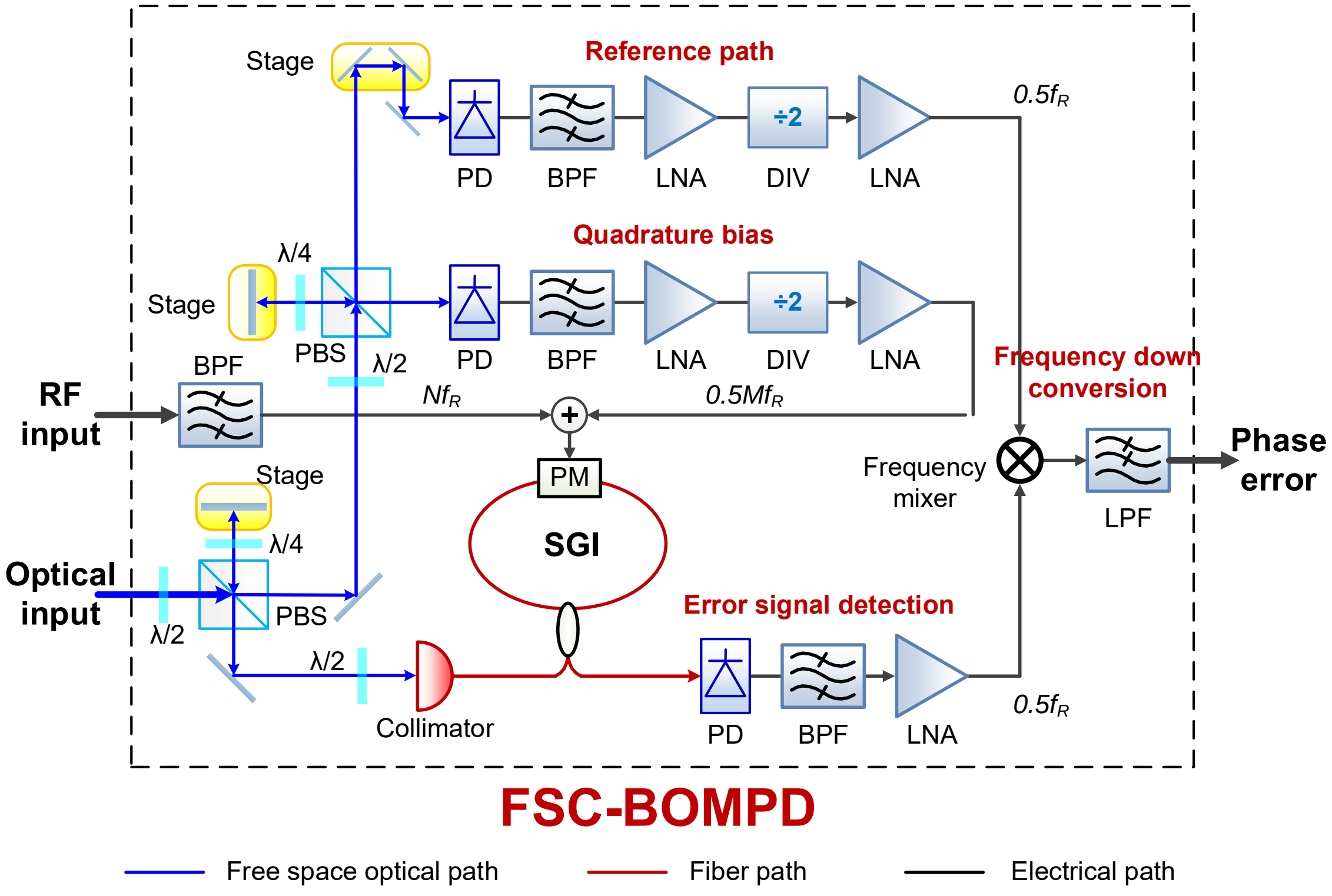

**Figure S7.** Detailed scheme of the free-space-coupled balanced optical-microwave phase detector (FSC-BOMPD) (BPF, bandpass filter; LPF, lowpass filter; PD, photodetector; LNA, low-noise amplifier; DIV, frequency divider; PM, phase modulator; $f_R$, the repetition rate of the optical input pulse train).

Compared with other BOMPD structures[S4], our FSC-BOMPD is insensitive to optical input power fluctuations. Besides that, phase fluctuations from the bias and reference paths also cannot affect the phase error output. So it can accurately measure the phase difference between the RF and optical input. An analytic derivation of the output voltage of the FSC-BOMPD is given below. To avoid ambiguity, all the symbols in the following equations are independent with those in Eqs. (S1-29). If the optical input pulse is sufficiently short, the optical pulse train power at the SGI input can be approximated by

$$P_{in}(t) = P_a T_R \sum_{n=-\infty}^{\infty} \left(1+\Delta_{RIN}(t)\right)\delta(t-nT_R-\Delta_J(t)) \tag{S30}$$

where $\delta(t)$ is a Dirac function, $T_R = 1/f_R$ is the period of the pulse train, $P_a$, $\Delta_{RIN}(t)$ and $\Delta_J(t)$ are the average power, power fluctuation and timing jitter of the pulse train, respectively.

The driving signal of the phase modulator can be written as

$$\varphi(t) = \Phi_0 \sin\left[2\pi f_0\left(t+\Delta t_0\right)\right] + \Phi_b \sin\left[2\pi(M+0.5)f_R\left(t+\Delta t_b\right)+\Delta\phi\right] \tag{S31}$$

where $\Phi_0$, $f_0$ and $\Delta t_0$ are the amplitude, frequency and timing jitter of the RF input signal, respectively. $\Phi_b$ is the amplitude of the RF signal from the quadrature bias path. $\Delta\phi$ and $\Delta t_b$ are the fixed relative phase and relative timing jitter between the pulse train and the RF bias signal, respectively. $M$ is an integer.

After circulating in the SGI, the output optical power can be expressed as

$$\begin{aligned} P(t) &= (1-\alpha)P_{in}(t)\sin^2(\varphi/2) \\ &= (1-\alpha)P_a T_R \times \\ &\sum_{n=-\infty}^{\infty} \sin^2\left[\frac{1}{2}\Phi_0 \sin\left[2\pi f_0 t+\theta_e\right]+\frac{1}{2}\Phi_b \sin\left[2\pi(M+0.5)f_R t+\theta_b+\Delta\phi\right]\right]\left(1+\Delta_{RIN}\right)\delta(t-nT_R) \end{aligned} \tag{S32}$$

where α is the loss of the SGI, $\theta_e=2\pi f_0(\Delta t_0+\Delta_J)$ is the relative phase error between the SGI and RF input signals, and $\theta_b=2\pi(M+0.5)f_R\Delta t_b$ is the phase fluctuation of the bias path.

Suppose the frequency of the RF and optical input signals are locked with each other by the FSC-BOMPD, then $f_0=Nf_R$, $N$ is an integer. Since $\delta(t-nT_R)$ is nonzero only if $t=nT_R$, and $\theta_e$, $\theta_b$ <<1, Eq. (S32) can be simplified as

$$P(t) \approx \frac{1}{2}(1-\alpha)P_a T_R \sum_{n=-\infty}^{\infty}\left[1-\cos\left(\Phi_0\theta_e+(-1)^n\Phi_b\sin(\Delta\phi+\theta_b)\right)\right]\left(1+\Delta_{RIN}\right)\delta(t-nT_R) \tag{S33}$$

Let

$$\Phi_b \sin(\Delta\phi) = \pi/2 \tag{S34}$$

Eq. (S33) can be written as

$$P(t)=\frac{1}{2}(1-\alpha)P_aT_R\sum_{n=-\infty}^{\infty}\left[1+\Phi_b\left(-\frac{1}{2}\sin\Delta\phi\theta_b^2+\cos\Delta\phi\sin\theta_b\right)\right]\left(1+\Delta_{RIN}\right)\delta(t-nT_R)$$
$$+\frac{1}{2}(1-\alpha)P_aT_R\sum_{n=-\infty}^{\infty}\Phi_0\theta_e\left(1+\Delta_{RIN}\right)\delta(t-nT_R)e^{j\pi f_R t} \tag{S35}$$

In the frequency domain, the first item on the right side of Eq. (S35) represents the components at $Kf_R$, while $(K+1/2)f_R$ for the second item ($K$ is an integer). Since $\theta_b$ only appears in the first item, the phase fluctuation from the bias path cannot affect the down-conversion signal at $f_R/2$.

Assume the RF output signal from the reference path is

$$V_r=\Phi_R\sin\left(\pi f_R(t+\Delta t_r)\right) \tag{S36}$$

where $\Phi_R$ is the amplitude of the reference signal, and $\Delta t_r$ is the relative timing jitter between the reference path and the SGI path. Then the error signal after down conversion is

$$V_e=\frac{C}{2}(1-\alpha)P_aT_RV_r\left(1+\Delta_{RIN}\right)\cos\left(\pi f_R\Delta t_r\right)\Phi_0\theta_e \tag{S37}$$

where $C$ is a constant coefficient related to the mixer, low-pass filters and other RF components in the down conversion path. If $f_R$=216 MHz, 3 mm length change (10 ps) from the reference path can only introduce $2\times10^{-5}$ change to $V_e$. So the phase fluctuation of the reference path contributes little to the output error signal. Similarly, since usually $\Delta_{RIN}$ <<1, the effect due to optical input power fluctuations $\Delta_{RIN}$ is also negligible. So $V_e$ is mainly determined by $\theta_e$, the relative timing jitter between the RF and SGI input signals. The free-space optical paths before the SGI are well isolated from environmental changes, so that the FSC-BOMPD can accurately detect the timing jitter between the RF and optical input signals without introducing systematic errors.

## V. Synchronous laser-microwave network

A detailed schematic of the laser-microwave network characterization setup is given in Fig. S8, that refers to Figs. S3, S5 and S7. A sapphire-loaded crystal oscillator (SLCO) serves as the remote VCO source. It operates at a 10.833-GHz center frequency, which is an integer multiple (×50) of the master laser repetition rate. A low-phase-noise amplifier (LPNA) after the SLCO is used to maximize the phase discriminant of the FSC-BOMPDs. In each FSC-BOMPD (see Fig. S7), the optical input is first separated into the SGI, bias and reference paths with free space PBSs. The pulse train in the bias path is detected, filtered at the 59$^{th}$ harmonic of the repetition rate, frequency-divided to 6.37 GHz (=29.5$f_R$), and adjusted in amplitude and phase to achieve Eq. (S34) for quadrature bias. Using similar detection electronics, the reference signal is generated at 108 MHz (=0.5$f_R$). When the input

RF signal is applied to SGI, any phase error between the input signals will induce amplitude modulation in the pulse train at the SGI output. The error modulation sidebands of the pulse train are directly detected, filtered at 108 MHz and down-converted in-phase with the 108-MHz reference signal to generate the output phase error signal at baseband. The phase error voltage signal in the VCO locking FSC-BOMPD is fed back to the frequency tuning port of the SLCO to achieve remote optical-to-microwave synchronization through the 1.2-km link. The out-of-loop FSC-BOMPD compares the relative timing error between the remote laser and the SLCO, to evaluate the performance of the whole synchronization network.

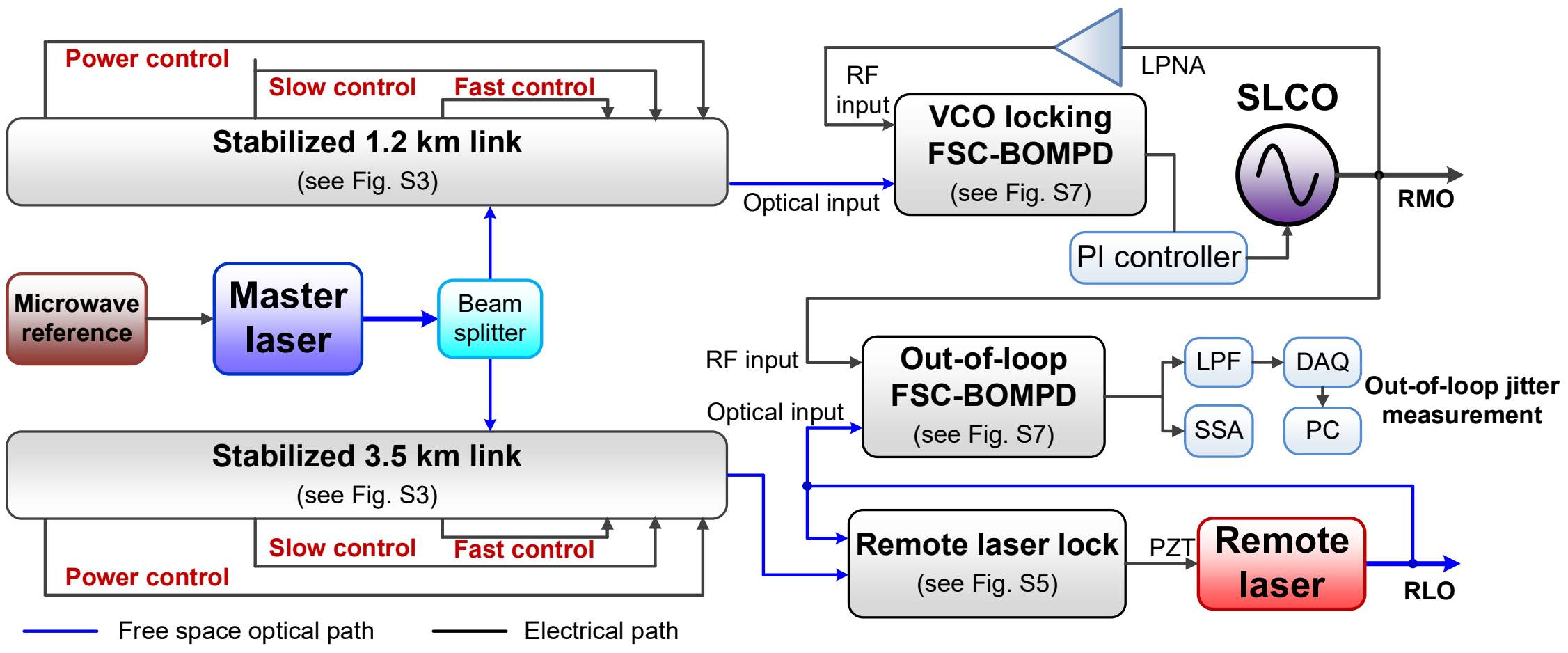

**Figure S8.** Detailed scheme of laser-microwave network synchronization.

To avoid fiber nonlinearities in the 1.2-km link, input optical power into the VCO locking FSC-BOMPD is about +9 dBm: both bias and reference arms utilize 0 dBm while the SGI obtains +5 dBm after collimator loss. For the out-of-loop FSC-BOMPD, due to all-free-space optics in the remote laser part, there is no nonlinearity onset and totally +14 dBm is provided to the optical input: 0 dBm for bias and reference paths and +10 dBm for SGI after collimator, respectively. The RF input power for each FSC-BOMPD is about +21 dBm. The phase sensitivity of the VCO locking and out-of-loop FSC-BOMPDs are about 0.25 mV/fs and 2.5 mV/fs, respectively. In Fig. 4g (black curve), The 20-kHz locking bandwidth is limited by the excitation of higher-order poles and zeroes due to high feedback gain. In some parts of the FSC-BOMPDs, 1 fs of the phase difference translates to a voltage shift of only several micro-volts. Therefore, all digital circuits are galvanically isolated from the FSC-BOMPDs, while the analog power lines are connected in a single point configuration through electromagnetic interference filters to eliminate undesirable ground loops and noise pick-ups and prevent electrostatic discharge. Thus, robust functioning of the FSC-BOMPD in actual service has been achieved without dropping out of lock during several weeks of operation. The FSC-BOMPDs have no active temperature control system; however, they are placed within a closed thermal insulated chamber to damp temperature and humidity changes of the environment. After the necessary warm-up time, the two FSC-BOMPDs do not experience more than 0.08 K of temperature drift and 0.6% relative humidity change during the measurement period (Fig. S9).

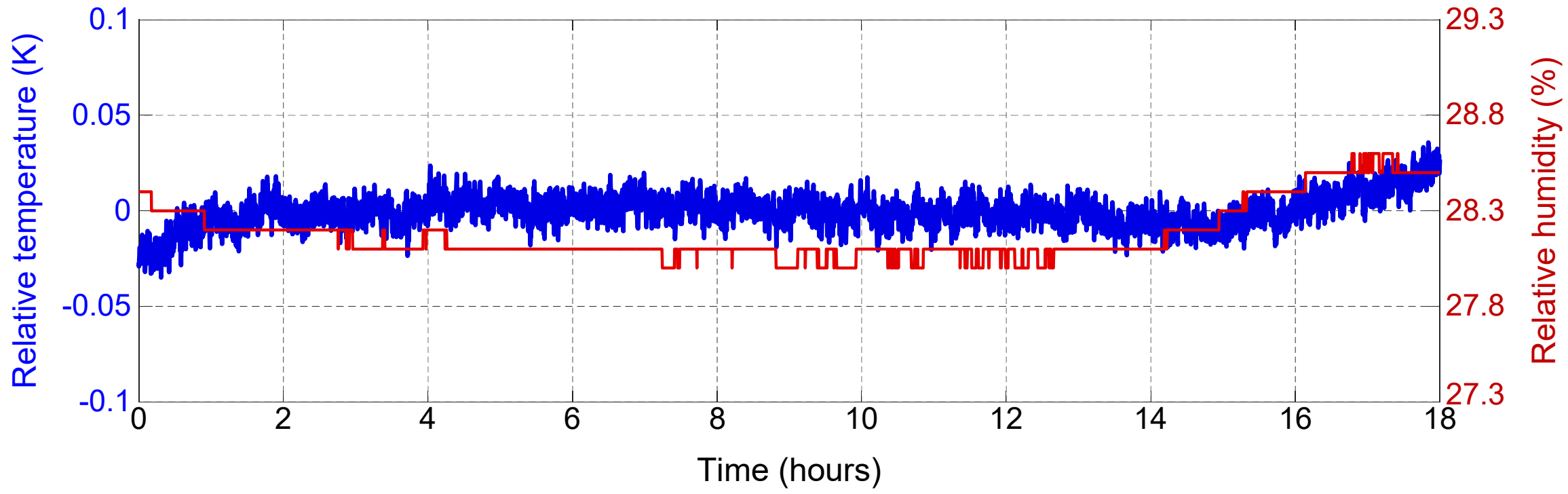

**Figure S9.** The environmental temperature and humidity change of the two FSC-BOMPDs in laser-microwave network characterization.